\documentclass{ws-ijmpcs}

\usepackage{hyperref}

\def\trimmarks{}

\begin{document}

\markboth{Schweitzer, Strikman, and Weiss}
{Sea quark transverse momentum distributions\ldots}

%\catchline{}{}{}{}{}
%
%%%%%%%%%%%%%%%%%%%%%%%%%%%%%%%%%%%%%%%%%%%%%%%%%%%%%%%%%%%%%%%%%%%%

\title{SEA QUARK TRANSVERSE MOMENTUM DISTRIBUTIONS \\[0ex] 
AND DYNAMICAL CHIRAL SYMMETRY BREAKING\footnote{Proceedings of 
QCD Evolution Workshop, Jefferson Lab, May 6--10, 2013,
\url{http://www.jlab.org/conferences/qcd2013/}}}

\author{P.~SCHWEITZER}
\address{Department of Physics, University of Connecticut,
Storrs, CT 06269, USA}

\author{M.~STRIKMAN}
\address{Department of Physics, Pennsylvania State University,
University Park, PA 16802, USA}

\author{C.~WEISS}
\address{Theory Center, Jefferson Lab, Newport News, VA 23606, USA}

\maketitle

%\begin{history}
%\received{Day Month Year}
%\revised{Day Month Year}
%\end{history}

\begin{abstract}
Recent theoretical studies have provided new insight into the intrinsic 
transverse momentum distributions of valence and sea quarks in the
nucleon at a low scale. The valence quark transverse momentum distributions 
($q - \bar q$) are governed by the nucleon's inverse hadronic 
size $R^{-1} \sim 0.2\, \textrm{GeV}$ and drop steeply at large $p_T$. 
The sea quark distributions ($\bar q$) are in large part generated by 
non--perturbative chiral--symmetry--breaking interactions and extend 
up to the scale $\rho^{-1} \sim 0.6 \, \textrm{GeV}$.
These findings have many implications for modeling the initial conditions
of perturbative QCD evolution of TMD distributions (starting scale, 
shape of $p_T$ distributions, coordinate--space correlation functions).
The qualitative difference between valence and sea quark intrinsic $p_T$
distributions could be observed experimentally, by comparing the transverse 
momentum distributions of selected hadrons in semi--inclusive 
deep--inelastic scattering, or those of dileptons produced in $pp$ and 
$\bar p p$ scattering.

\end{abstract}

\ccode{PACS numbers: 11.15.Pg, 12.38.Lg, 12.39.Fe, 13.60.Hb, 
13.87.-a, 13.88.+e. \\ Report number: JLAB-THY-13-1785}
\section{Transverse momentum distributions in QCD}	
Describing the transverse momentum distributions of particles
produced in hard processes in high--energy $eN$ and $pp/\bar p p$ scattering
(semi--inclusive deep--inelastic scattering or DIS, 
Drell--Yan pair production) has been a focus of recent theoretical 
research in Quantum Chromodynamics (QCD). At sufficiently large 
transverse momenta 
$P_T \sim \textrm{few GeV}$ the observed particle distributions 
are generated by individual QCD processes and can be computed in
fixed--order perturbation theory, starting from the 
well--known collinear (i.e., integrated over transverse momenta) 
parton distributions in the initial nucleon(s). The scale dependence 
of these functions due to QCD radiation is described by the Dokshitzer 
et al.\ (DGLAP) evolution equations. At lower transverse momenta the 
observed $P_T$ distributions are the result of an interplay of 
several factors: the intrinsic transverse momentum of the partons 
in the nucleon, soft QCD final--state interactions, and the transverse 
momentum incurred in the parton fragmentation process. QCD radiation 
in this kinematics is subject to Sudakov suppression and leads to
evolution equations of Collins--Soper--Sterman (CSS) type for the
transverse momentum distributions.\cite{Collins:1984kg}
Considerable progress has been made\cite{Collins:2013zsa}
in formulating a factorized description
of semi--inclusive DIS at low $P_T$, establishing the QCD operator 
definitions of the pertinent transverse momentum dependent (or TMD) 
parton distribution and fragmentation functions, and deriving the 
CSS--type QCD evolution equations for the 
latter.\cite{Aybat:2011zv,Sun:2013hua} In order to apply 
this formalism to actual data one needs to understand the basic 
properties of the TMD distributions at a low scale, which represent
the initial condition for the solution of the evolution equations,
as determined by non--perturbative QCD interactions. This includes
the dynamical mechanisms producing intrinsic transverse 
momentum in the nucleon, the shape of the distributions, and the natural 
starting scale for perturbative QCD evolution.\cite{Schweitzer:2012hh} 
\section{Valence and sea quark transverse momentum distributions}
\label{sec:valence_sea}
Of particular interest is a comparison of the transverse momentum 
distributions of ``valence'' quarks, $f_1^{q - \bar q}(x, p_T)$, and 
``sea'' quarks $f_1^{\bar q}(x, p_T)$, at a low 
scale.\cite{Schweitzer:2012hh} Since they
are created by different non--perturbative mechanisms one expects 
these distributions to have different properties. The essential 
points can be explained with heuristic arguments, to be supported
by dynamical model calculations later.

The nucleon's valence quark structure generally follows the pattern of 
a bound state with fixed particle number and approximately
independent motion of the constituents (mean--field picture).
Model--independent evidence for the approximate mean--field character
of the nucleon's valence quark light--cone wave function comes \textit{e.g.} 
from an analysis of the empirical transverse charge densities, which 
shows that the $u/d$ ratio of densities is practically constant over 
a wide range of distances $b \lesssim 1\, \textrm{fm}$.\cite{Miller:2011du} 
In such mean--field systems the single--particle 
momentum--space wave functions are Fourier--conjugate to the corresponding
coordinate--space functions. Up to trivial effects of relativistic 
kinematics the transverse momentum distributions are therefore governed
by a single dynamical scale, namely the inverse overall size of the 
bound state, $R^{-1} \sim (1\, \textrm{fm})^{-1} = 0.2\, \textrm{GeV}$. 
Such behavior is indeed observed in a variety
of relativistic quark models based on the mean--field approximation
(bag model, covariant bound--state models, light--front 
models).\cite{Schweitzer:2010tt}

\begin{figure}[t]
\begin{tabular}{ll}
\includegraphics[width=4.3cm]{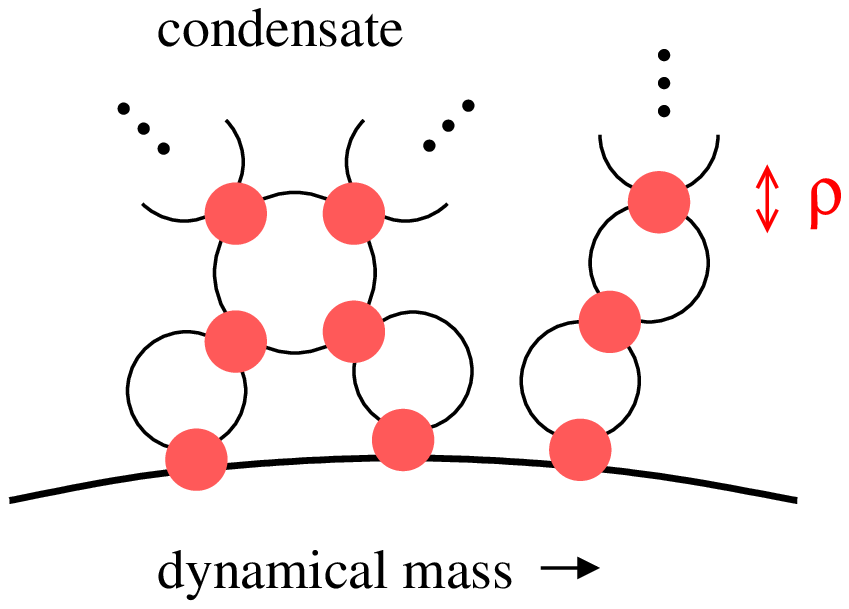}
\hspace{1.3cm}
&
\includegraphics[width=5.3cm]{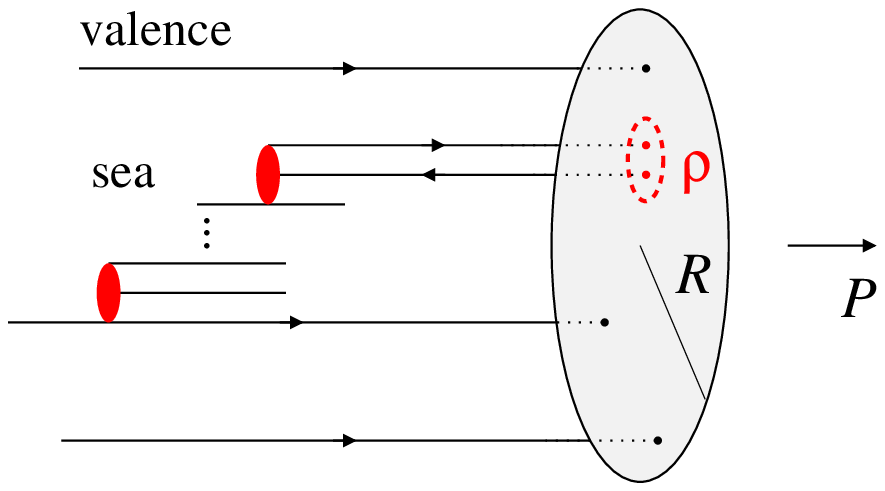}
\\[-1ex]
{\scriptsize (a)} & {\scriptsize (b)}
\end{tabular}
\vspace*{8pt}
\caption[]{(a) Dynamical chiral symmetry breaking in QCD. The red
blobs indicate non--perturbative interactions with a characteristic 
range $\rho \sim 0.3\, \textrm{fm}$.
(b) Schematic illustration of the dynamical scales governing
the valence and sea quark transverse momentum distributions
at a low scale.\cite{Schweitzer:2012hh}}
\label{fig:chisb}
\end{figure}

Sea quarks in the nucleon's light--cone wave functions appear due to 
non-perturbative interactions at distance scales generally unrelated 
to the overall nucleon size, including much shorter distances.
Of particular importance are the short--range forces responsible for 
the dynamical breaking of chiral symmetry in QCD. In the Euclidean
(imaginary--time) formulation of QCD such forces are induced by 
topologically charged gauge fields of characteristic size 
$\rho \sim 0.3 \, \textrm{fm} \ll R$, whose properties have been studied 
extensively in lattice simulations and analytic approximation
schemes.\cite{Diakonov:2002fq} 
These interactions are responsible for the appearance of a condensate
of quark--antiquark pairs in the vacuum (see Fig.~\ref{fig:chisb}a), 
and, more generally, 
for the dynamical generation of the masses of light hadrons in QCD.
There is strong evidence that a large part of the quark--antiquark sea
in the nucleon's partonic structure is due to such chiral--symmetry--breaking
interactions; \textit{e.g.}\ in the observed non--trivial flavor structure
of the sea. This would imply that sea quarks can have transverse momenta
of the order of the chiral--symmetry--breaking scale 
$\rho^{-1} \sim 0.6 \, \textrm{GeV}$, much larger than the inverse 
nucleon size $R^{-1} \sim 0.2\, \textrm{GeV}$ 
(see Fig.~\ref{fig:chisb}b). One would thus expect
the $p_T$ distribution of sea quarks to be qualitatively different
from that of valence quarks.

In the terminology of nuclear physics, the nucleon in QCD
represents a many--body system with short--range correlations induced by 
the chiral--symmetry--breaking interactions. The transverse structure is 
determined by \textit{two dynamical scales:} the overall size of the 
system, $R$, and the size of the correlations, $\rho \ll R$. The valence 
and sea quark transverse momentum distributions are affected by 
these dynamical scales in different ways. This basic feature is
principally not described by single--scale mean--field models
of the nucleon.
\section{Dynamical model based on chiral symmetry breaking}
\begin{figure}
\centerline{\includegraphics[width=10cm]{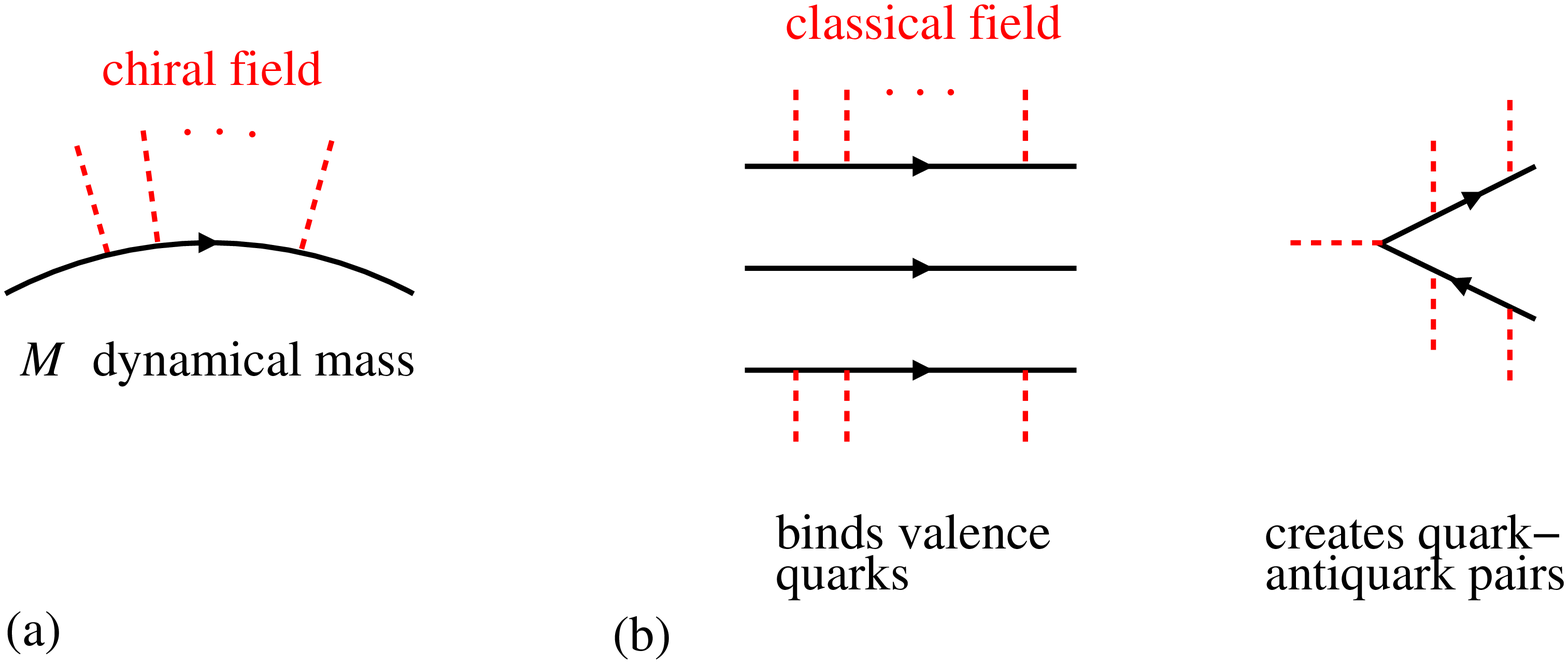}}
\caption[]{Chiral quark--soliton model of the nucleon. (a) Effective dynamics.
The quarks acquire a dynamical mass, accompanied by the coupling to 
a chiral field. (b) Nucleon solution in the large--$N_c$ limit. 
The classical chiral field (soliton) binds the valence quarks and 
creates quark--antiquark pairs.}
\label{fig:chiraleff}
\end{figure}

The qualitative difference between the valence and sea quark transverse 
momentum distributions can be illustrated\cite{Schweitzer:2012hh} 
in a model of the nucleon
that implements the effective low--energy dynamics resulting from chiral 
symmetry--breaking in QCD.\cite{Diakonov:1987ty} It uses 
``constituent'' quarks/antiquarks with a dynamical mass 
$M \sim 0.3-0.4 \, \textrm{GeV}$ as effective degrees of freedom below 
the chiral symmetry--breaking scale. The dynamical mass is accompanied 
by a coupling to a chiral field describing the long--wavelength phase 
fluctuations of the chiral condensate (Goldstone bosons, 
see Fig.~\ref{fig:chiraleff}a). The effective
coupling constant is $M/F_\pi \sim$ 3--4, so that the dynamical system
is strongly coupled and has to be solved non--perturbatively 
using the $1/N_c$ expansion. The effective dynamics applies to momenta
up to the chiral symmetry breaking scale $\rho^{-2}$, which acts
as an ultraviolet cutoff of the model.

%
% FIGURE
%
The nucleon in the effective model develops a classical chiral field;
in the rest frame it is of a generalized spherical form (``hedgehog'') 
and has a characteristic radius $R \sim M^{-1}$. The classical 
chiral field acts in a dual way: it binds $N_c$ valence quarks in a
discrete bound--state level and distorts the chiral condensate, 
amounting to the coherent creation of additional quark--antiquark pairs 
out of the vacuum (chiral quark--soliton model, or
relativistic mean--field approximation, see 
Fig.~\ref{fig:chiraleff}b).\cite{Diakonov:1987ty} 
Because the dynamics is formulated as a field theory it guarantees
completeness of the quark single--particle states and preserves the
partonic sum rules and positivity conditions of QCD, in the sense of 
a parametric expansion based on the hierarchy $\rho^{-2} \gg M^2$. 
The model is therefore uniquely suited to describe the nucleon's parton 
densities at a low scale, especially sea quark distributions.

The calculation of parton distributions in the chiral quark--soliton
model has been described in detail in the 
literature.\cite{Diakonov:1996sr,Diakonov:1997vc} The quark and
antiquark densities can be computed either as number densities
of field quanta in the infinite--momentum frame,\footnote{It was 
recently proposed that a similar 
approach could be used to calculate the QCD 
parton densities directly as functions of $x$, expressing them
as Euclidean correlation functions that could be computed in
lattice QCD.\cite{Ji:2013dva}} 
or as light--cone correlation functions of the fields
in the rest frame; the two formulations are equivalent thanks to
the relativistic invariance and completeness of the model 
dynamics.\cite{Diakonov:1997vc}
The transverse momentum integrals extend up to values of the
order of the chiral--symmetry breaking scale $\rho^{-2}$,
so that the model describes the ``intrinsic'' transverse momentum 
distributions due to non--perturbative nucleon structure.
The model does not include effects of final--state interactions.

\begin{figure}
\begin{tabular}{ll}
\includegraphics[width=4.2cm]{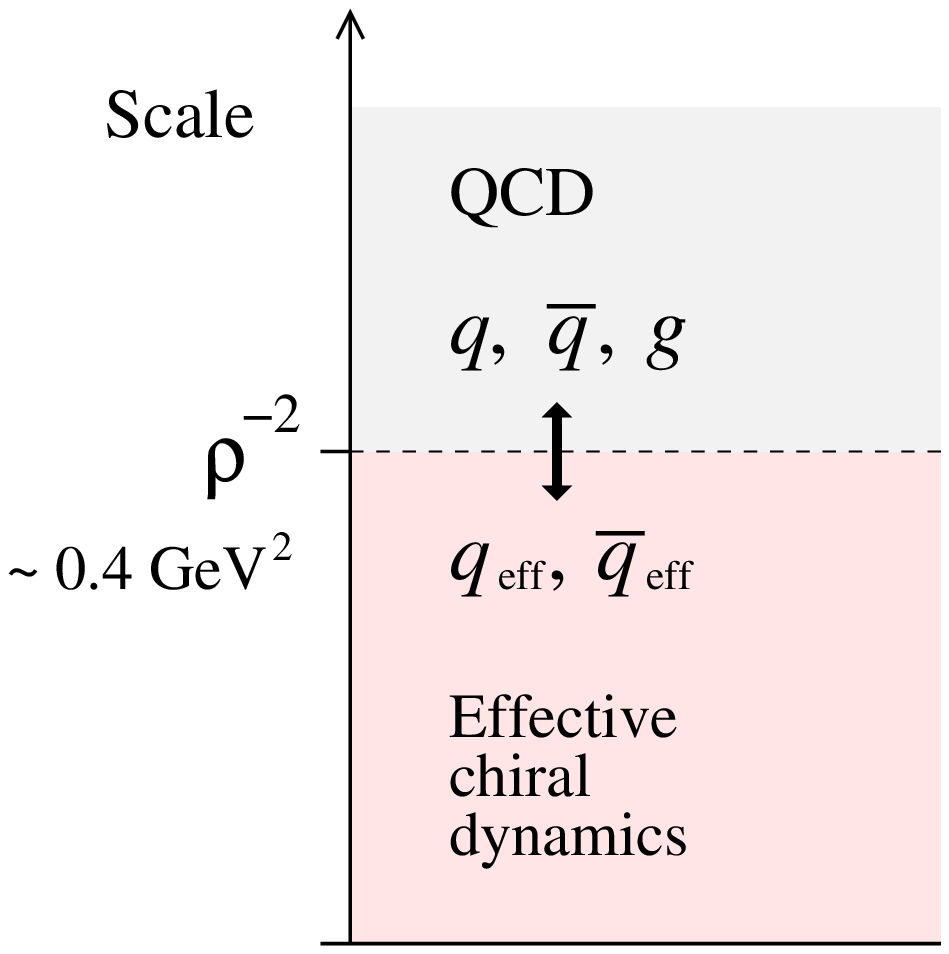}
\hspace{1cm}
&
\includegraphics[width=7cm]{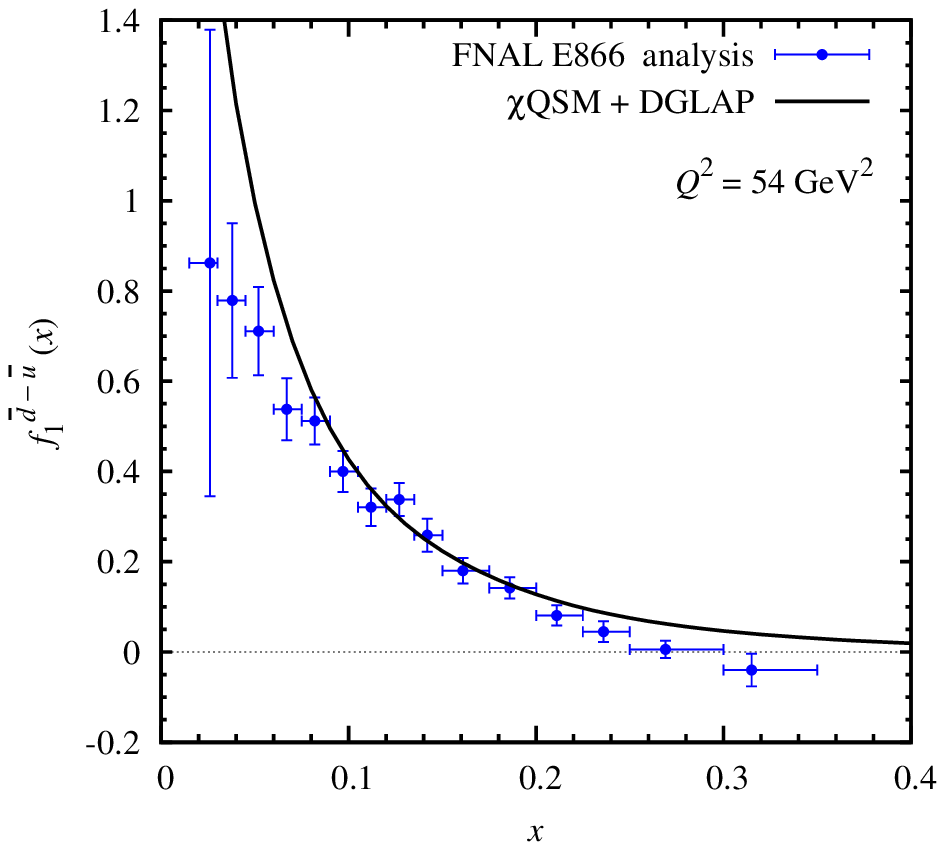}
\end{tabular}
\caption[]{(a)~Matching of the effective chiral model with QCD at the
chiral symmetry--breaking scale $\rho^{-2}$. (b)~Flavor--nonsinglet
sea quark density in the chiral quark--soliton 
model,\cite{Pobylitsa:1998tk} after DGLAP
evolution from the scale $\rho^{-2}$ to $Q^2 = 54\,\textrm{GeV}^2$, 
compared to the FNAL E866 Drell--Yan analysis.\cite{Towell:2001nh}}
\label{fig:match}
\end{figure}

The parton distributions calculated in the chiral quark--soliton model
are the light--cone momentum distributions of effective degrees
of freedom --- constituent quarks and antiquarks, which are to be
matched with QCD quarks, antiquarks and gluons at the chiral
symmetry--breaking scale $\rho^{-2}$ (see Fig.~\ref{fig:match}a).
The matching can be performed either on the basis of a ``microscopic''
derivation of the effective chiral model from QCD, such as the
instanton vacuum model,\cite{Diakonov:1985eg,Diakonov:1995qy}
or with the help of empirical parton densities obtained from fits 
to DIS data. In the simplest approximation the composite quarks 
and antiquarks are identified with the QCD quarks and antiquarks
at the scale $\rho^{-2}$, and the gluon density is set to zero.
Its accuracy can be judged from the fact that in empirical 
leading--order parton densities\cite{Gluck:2007ck} at the scale
$\mu_{\rm LO}^2 = 0.3 \, \textrm{GeV}^2$
about $\sim 30\%$ of the nucleon's light--cone momentum is carried 
by gluons. Of particular significance is that the model describes
well\cite{Pobylitsa:1998tk} the observed flavor--nonsinglet unpolarized sea 
$f_1^{\bar d - \bar u}(x) \equiv \bar d(x) - \bar u(x)$,\cite{Towell:2001nh} 
which is expected to be much less sensitive to ``matching'' effects than
the flavor--singlet distributions (see Fig.~\ref{fig:match}b).
The model also predicts a large flavor--nonsinglet polarized 
sea,\cite{Diakonov:1996sr,Diakonov:1997vc} hints of which are seen 
in recent global QCD fits including 
semi--inclusive DIS data and the RHIC $W^\pm$ production 
data.\cite{Surrow}
%
% FIGURE
%

\begin{figure}[t]
\begin{tabular}{ll}
\psfig{file=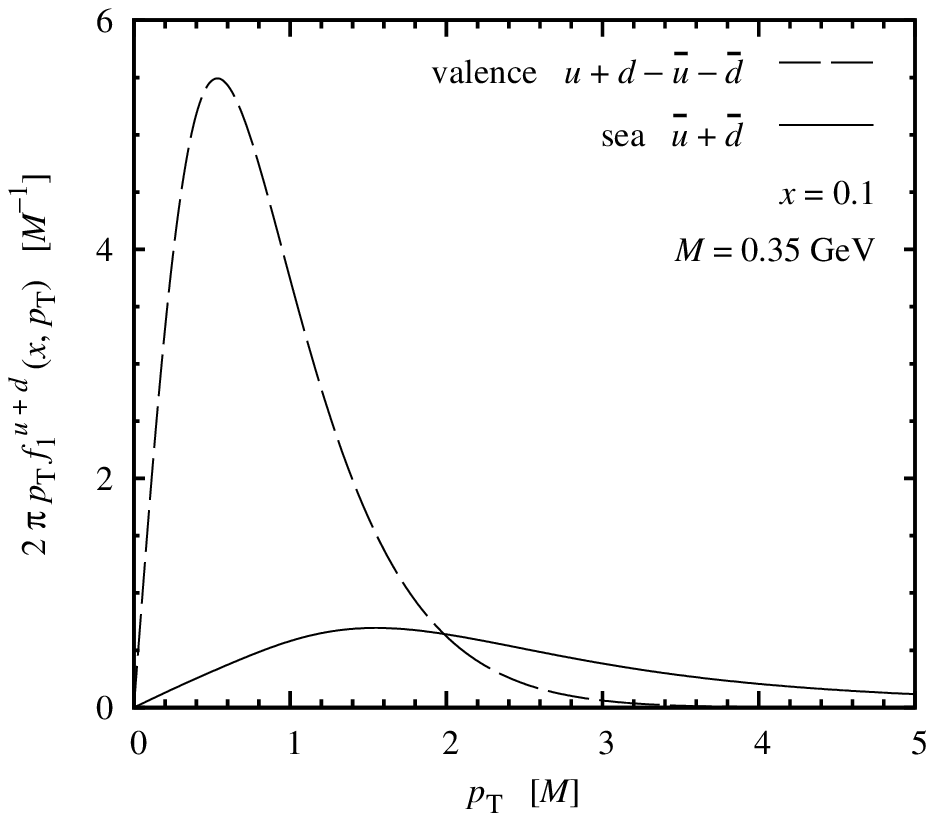,width=6.7cm}
&
\hspace{-2em}
\psfig{file=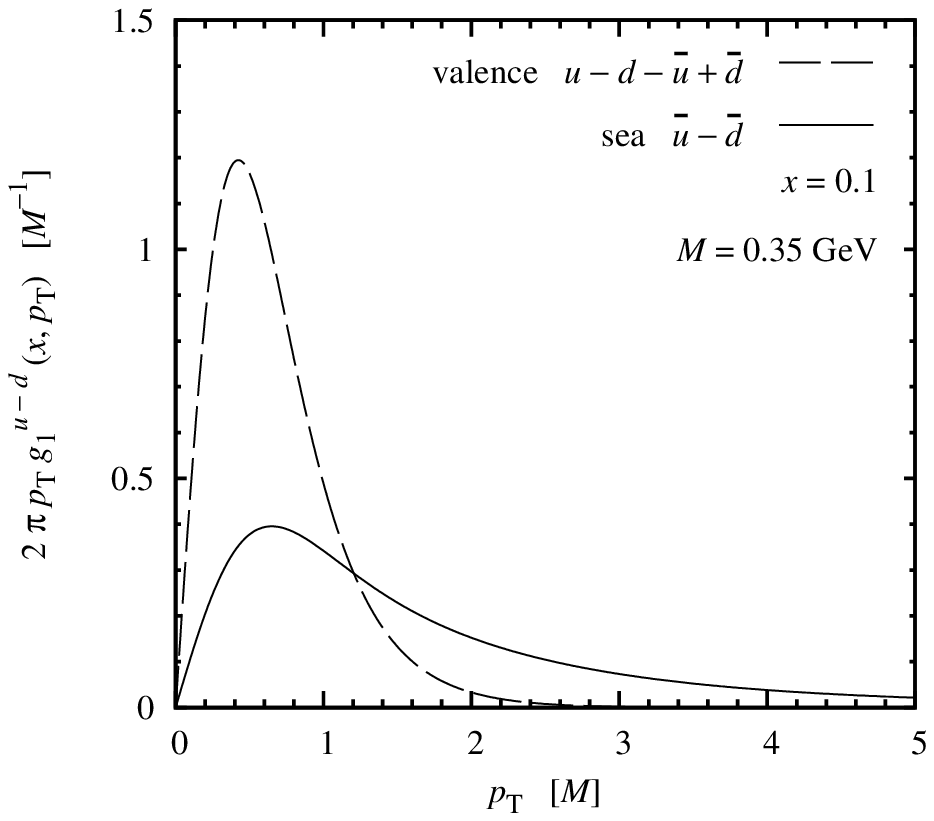,width=6.7cm}
\\[-2ex]
{\scriptsize (a)} & {\scriptsize (b)} 
\end{tabular}
%\vspace*{8pt}
\caption[]{(a) Intrinsic transverse momentum distributions of flavor--singlet
unpolarized valence quarks $f_1^{u + d - \bar u - \bar d}$ (dashed line) 
and sea quarks $f_1^{\bar u + \bar d}$ (solid line) at $x = 0.1$ in the chiral 
quark--soliton model.\cite{Schweitzer:2012hh} The plot shows the radial
distributions $2\pi p_T f_1(x, p_T)$, so that the area under the curve
corresponds to the $p_T$--integrated parton density in the model. 
The distributions and $p_T$ are given 
in units of the constituent quark mass $M = 0.35\, \textrm{GeV}$.
(b) Distributions of flavor--nonsinglet polarized valence quarks
$g_1^{u - d - \bar u + \bar d}$ (dashed line) and sea quarks 
$g_1^{\bar u - \bar d}$ (solid line).}
\label{fig:f1_val_sea}
\end{figure}

Figure~\ref{fig:f1_val_sea}a shows the intrinsic $p_T$ distributions
of flavor--singlet unpolarized quarks at $x = 0.1$ in the 
chiral quark--soliton model.\cite{Schweitzer:2012hh} 
The valence quark distribution $f_1^{u + d - \bar u - \bar d}(x, p_T)$ 
drops steeply with increasing $p_T$ and can roughly be approximated
by a Gaussian shape. The average transverse momentum of the valence quarks 
is of the order $\langle p_T^2 \rangle \sim M^2 \sim R^{-2}$, 
corresponding to the inverse radius of the mean field binding the valence 
quarks (cf.\ Sec.~\ref{sec:valence_sea} and Fig.~\ref{fig:chisb}b).
The sea quark distribution $f_1^{\bar u + \bar d}(x, p_T)$ extends up
to much larger values of $p_T$. Closer inspection of the analytic
expressions shows that it contains a ``would--be'' power--like tail of
the form $f_1^{\bar u + \bar d}(x, p_T) \sim C(x)/(p_T^2 + M^2)$,
which is regulated by the UV cutoff representing the chiral symmetry--breaking
scale $\rho^{-1}$. The coefficient $C(x)$ is determined by low--energy chiral 
dynamics at momenta of the order $M$ and model--independent.
At $p_T \gg M$ the distributions exhibit some residual
model dependence, due to choice of cutoff scheme implementing the
chiral symmetry--breaking scale, which was studied 
numerically\cite{Schweitzer:2012hh} and found to be minor up to 
$p_T \sim 3\, M \sim 1\, \textrm{GeV}$.
These results clearly illustrate the qualitative difference between 
the valence and sea quark $p_T$ distributions due to dynamical chiral
symmetry breaking. The features described here do not depend on the 
details of the model but rely only on the existence of two separate 
dynamical scales --- the nucleon size $R$, and the chiral 
symmetry--breaking scale $\rho$.

Similar behavior is found\cite{Schweitzer:2012hh} 
in the transverse momentum distributions
of flavor--nonsinglet polarized quarks, 
$g_1^{u - d - \bar u + \bar d}(x, p_T)$ 
and $g_1^{\bar u - \bar d}(x, p_T)$ 
(see Fig.~\ref{fig:f1_val_sea}b), which
appear in the same order of the $1/N_c$ expansion as the flavor--singlet
unpolarized distributions and share many features with 
them.\cite{Diakonov:1996sr,Diakonov:1997vc} 
The flavor--nonsinglet polarized sea quark distribution 
$g_1^{\bar u - \bar d}(x, p_T)$ exhibits a would--be power--like tail
of a form analogous to that of the flavor--singlet unpolarized distribution.
The flavor--nonsinglet polarized valence quark distribution 
$g_1^{u - d - \bar u + \bar d}(x, p_T)$ 
drops more rapidly at large $p_T$ than the flavor--singlet unpolarized one,
making the discrepancy between sea and valence distributions at large
$p_T \sim 1\, \textrm{GeV}$ even more pronounced than in the unpolarized case.
We also note\cite{Schweitzer:2012hh} 
that the unpolarized and polarized $p_T$ distributions 
in this model obey a general inequality (positivity condition) and 
reflect in a subtle manner the restoration of chiral symmetry 
at $p_T \sim \rho^{-1}$. An important practical 
question is how the ``anomalously'' large intrinsic $p_T$ of 
the flavor--singlet polarized sea affect the analysis of semi--inclusive
DIS and $W^\pm$ production experiments aimed at extracting 
$g_1^{\bar u - \bar d}(x) \equiv \Delta\bar u(x) - \Delta\bar d(x)$,
if such measurements are performed with a finite acceptance in $p_T$.

The chiral quark--soliton model also permits to evaluate the coordinate--space
correlation functions,\cite{Schweitzer:2012hh} the Fourier 
transforms of the TMD distributions entering in the 
coordinate--space CSS evolution 
equations.\cite{Collins:2013zsa,Aybat:2011zv,Sun:2013hua} Because the
effective dynamics has a mass gap --- the constituent quark mass $M$,
the coordinate--space sea quark correlation function exhibits 
exponential behavior $\sim e^{-M \xi_T}$ over an intermediate range of 
distances $\rho \ll \xi_T \ll R$. At distances $\xi_T \sim R$ 
this behavior is modified by the spatial variation of the mean field;
because $R \sim M^{-1}$ the window for a visible exponential 
dependence is actually rather small.
\section{Experimental tests}
\begin{figure}[t]
\begin{tabular}{ll}
\includegraphics[width=5.8cm]{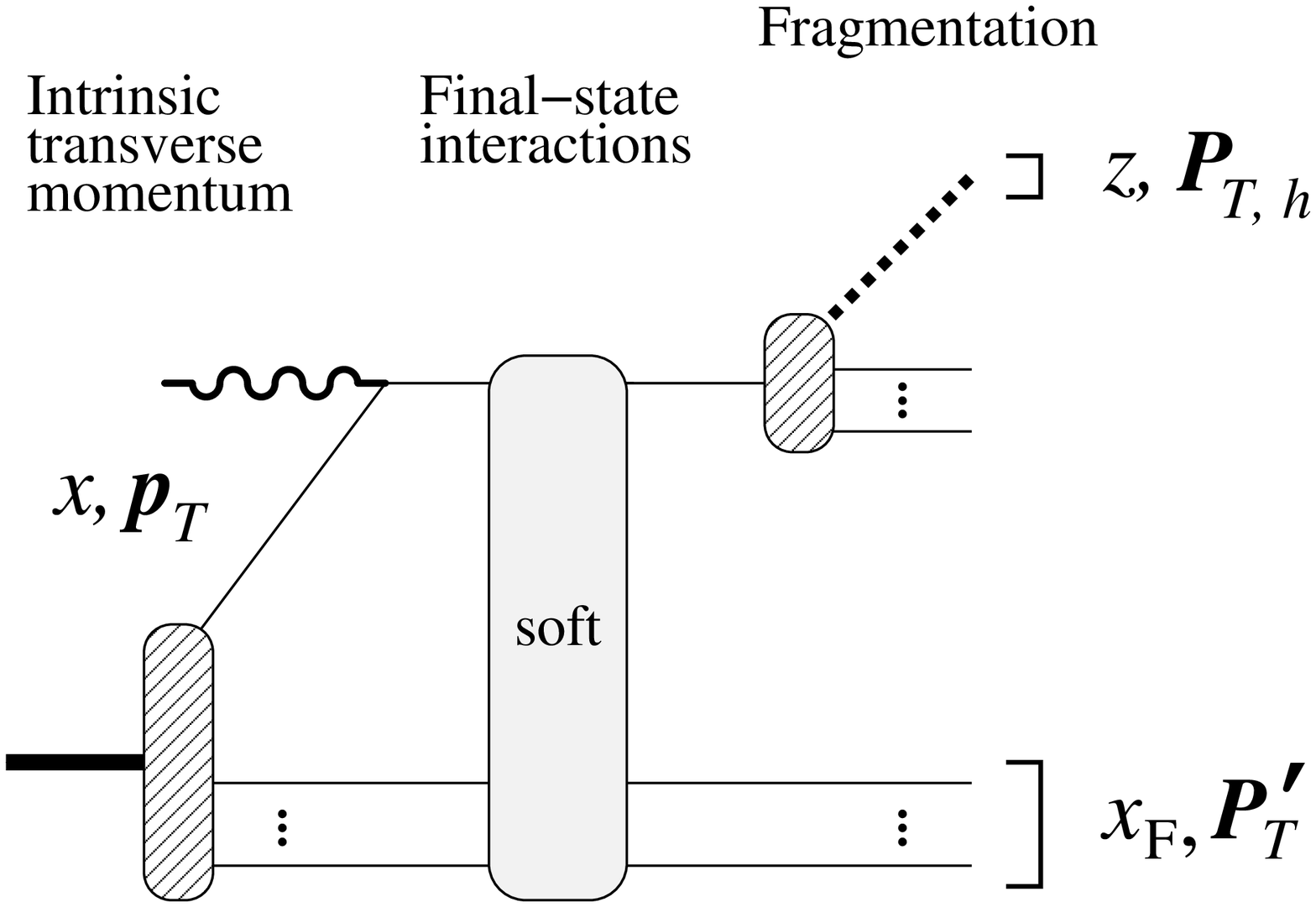}
\hspace{.5cm}
&
\includegraphics[width=6cm]{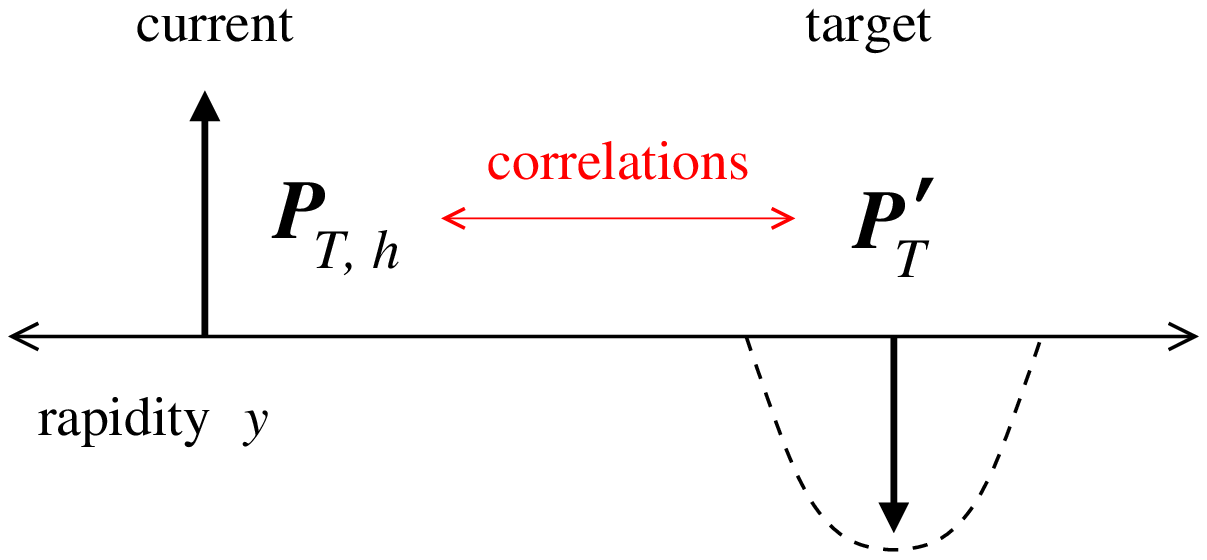}
\\[-1ex]
{\scriptsize (a)} & {\scriptsize (b)}
\end{tabular}
\vspace*{8pt}
\caption[]{(a) Semi--inclusive DIS in QCD. The transverse momentum 
$P_{T, h}$  of hadrons in the current fragmentation region is compounded 
from the intrinsic $p_T$ of the parton, soft final--state interactions, 
and the transverse momentum of fragmentation process. (b) Hadron 
correlations between the current and target fragmentation regions.}
\label{fig:sidis}
\end{figure}
The qualitative difference between the valence and sea quark $p_T$ 
distributions has numerous implications for semi--inclusive DIS 
measurements and could potentially be tested directly using 
special observables. In semi--inclusive DIS at 
$P_{T, h} \lesssim 1\, \textrm{GeV}$ the transverse momentum of the
identified hadron $h$ is compounded from the intrinsic $p_T$ of the 
struck parton, soft QCD final--state interactions, and the transverse
momentum incurred in the fragmentation process (see Fig.~\ref{fig:sidis}a).
The strength of the different mechanisms is poorly known at present,
making it difficult to quantify how differences in the intrinsic $p_T$
distributions express themselves in the observable hadron distributions.
Differential measurements of kinematic dependencies (e.g. $z$ distributions 
for fixed $x$, $P_{T, h}$ distributions for fixed $z$) could help to
disentangle the different mechanisms but require wide kinematic coverage
and high statistics. Detailed measurements of multiplicities have 
recently been reported by the HERMES and COMPASS 
experiments.\cite{Airapetian:2012ki,Adolph:2013stb} The valence quark 
region will be covered with high precision with the JLab 12 GeV Upgrade, 
but the kinematics is marginal for applying QCD factorization. A much 
broader kinematic region would become accessible with a future 
Electron--Ion Collider (EIC), permitting detailed studies of the production
mechanism, $Q^2$--evolution, and sea quark 
distributions.\cite{Accardi:2011mz,Accardi:2012hwp}

Special observables can detect systematic differences between the 
valence and sea quark $p_T$ distributions without detailed modeling
of the semi--inclusive production mechanism.\cite{Schweitzer:2012hh}
One possibility is to measure the difference and sum of the charged 
pion multiplicity distributions for a deuteron target (isoscalar), 
which in schematic notation are proportional to
\begin{eqnarray}
N_{\pi +, \; {\rm deut}} \; - \; N_{\pi -, \; {\rm deut}}\; 
& \propto & \; (e_u^2 - e_d^2) \; 
f_1^{u + d - \bar u - \bar d} \; \otimes 
\; D_1^{-} ,
\\[-.2ex]
N_{\pi +, \; {\rm deut}} \; + \; N_{\pi -, \; {\rm deut}}\; 
& \propto & \; (e_u^2 + e_d^2) \; 
f_1^{u + d + \bar u + \bar d} \; \otimes \;
D_1^{+} .
\end{eqnarray}
Here $f_1^{q \mp \bar q}$ denote the difference/sum of the quark and 
antiquark distributions in the proton and 
$D_1^{\mp} = D_1^{u/\pi +} \mp D_1^{\bar u/\pi +}$ the difference/sum
of the favored and unfavored pion fragmentation functions; isospin symmetry is
used and the contribution of the strange sea has been neglected. A broader
intrinsic $p_T$ distribution of sea quarks than of valence quarks should
generally manifest itself in a decrease of the ratio 
$(N_{\pi +, \; {\rm deut}} - N_{\pi -, \; {\rm deut}})/
(N_{\pi +, \; {\rm deut}} + N_{\pi +, \; {\rm deut}})$ with increasing
pion transverse momentum $P_{T, \pi}$,
or could simply be observed by comparing the normalized $P_{T, \pi}$
dependence of the difference and sum. Such measurements should be
performed at moderately small values $x \sim 0.1$, where valence and sea
quark densities are of comparable magnitude. Another possibility is to 
separate quarks and antiquarks in the target using charged kaons. 
$K^+$ are produced by favored fragmentation of $u$ quarks, whose 
distribution has both a valence and a sea component, while $K^-$ are 
produced from $\bar u$, which occurs in the sea only. Assuming identical 
transverse momentum distributions of strange quarks and antiquarks in the 
nucleon, $f_1^{s} = f_1^{\bar s}$, and neglecting differences in unfavored 
fragmentation, one expects the $K^-$ to have a broader $P_{T, K}$ 
distribution than the $K^+$.

An alternative experimental test of the different intrinsic $p_T$
distributions of valence and sea quarks would be comparing the
transverse momentum distributions of dileptons (Drell--Yan pairs)
produced in $pp$ and $\bar p p$ collisions. The pairs are produced
in the annihilation of a quark and and antiquark from the two colliding
hadrons. In $\bar p p$ collisions this is possible with valence quarks
and antiquarks, while in $p p$ the sea is involved in at least one 
of the protons. We therefore expect a broader dilepton $P_{T, \, l+l-}$
distribution in $pp$ than in $\bar p p$ in the same kinematics.
Again, such measurements would be most instructive at quark/antiquark
momentum fractions $x_{1, 2} \sim 0.1$, where valence and sea
quark densities are of comparable magnitude.

Much more insight could be gained from measurements of hadron correlations
between the current and the target fragmentation regions of semi--inclusive
DIS (see Fig.~\ref{fig:sidis}b).\cite{Schweitzer:2012hh} 
Such measurements could unravel the semi--inclusive production mechanism
by answering the question what ``balances'' the observed $P_{T, h}$ of hadrons
in the current fragmentation region --- other current fragments, 
central rapidity hadrons, or target fragments. They could discriminate
between scattering from sea and valence quarks by providing information
on the hadronic products of the remnant system (charge, flavor, 
multiplicities). Under certain conditions they could even reveal the
non--perturbative short--range correlations between sea quarks induced
by chiral symmetry breaking. This would require a rapidity
interval $\Delta y \gtrsim 4$ for clean separation of the current
and target regions, moderate scales $Q^2 \sim \textrm{few GeV}^2$
to avoid pQCD radiation, and access to quark/antiquark momentum fractions
$x \sim 0.1$ where the non--perturbative sea is large; these conditions
could be met in a ``window'' of moderate $\gamma^\ast N$ center--of--mass 
energies $W^2 \sim 30 \, \textrm{GeV}^2$. Such measurements could
ideally be performed with a medium--energy EIC with 
appropriate forward hadron detection capabilities.\cite{Accardi:2011mz} 
\\[2ex]
{\bf Notice:} Authored by Jefferson Science Associates, 
LLC under U.S.\ DOE Contract No.~DE-AC05-06OR23177. The U.S.\ Government 
retains a non--exclusive, paid--up, irrevocable, world--wide license to 
publish or reproduce this manuscript for U.S.\ Government purposes.

\begin{thebibliography}{00}
%
%
\bibitem{Collins:1984kg}
  J.~C.~Collins, D.~E.~Soper and G.~Sterman,
  %``Transverse Momentum Distribution In Drell-Yan Pair And W And Z Boson
  %Production,''
  Nucl.\ Phys.\ B {\bf 250}, 199 (1985).
  %%CITATION = NUPHA,B250,199;%%
%
%
\bibitem{Collins:2013zsa} 
  For a recent review, see: J.~Collins,
  %``TMD theory, factorization and evolution,''
  these proceedings, arXiv:1307.2920 [hep-ph].
  %%CITATION = ARXIV:1307.2920;%%
  %1 citations counted in INSPIRE as of 18 Aug 2013
%
%
\bibitem{Aybat:2011zv} 
  S.~M.~Aybat and T.~C.~Rogers,
  %``TMD Parton Distribution and Fragmentation Functions with QCD Evolution,''
  Phys.\ Rev.\ D {\bf 83}, 114042 (2011).
  %[arXiv:1101.5057 [hep-ph]].
  %%CITATION = ARXIV:1101.5057;%%
%
%
\bibitem{Sun:2013hua}
  P.~Sun and F.~Yuan,
  %``TMD Evolution: Matching SIDIS to Drell-Yan and W/Z Boson Production,''
  arXiv:1308.5003 [hep-ph].
  %%CITATION = ARXIV:1308.5003;%%
%
%
\bibitem{Schweitzer:2012hh} 
  P.~Schweitzer, M.~Strikman and C.~Weiss,
  %``Intrinsic transverse momentum and parton correlations from 
  %dynamical chiral symmetry breaking,''
  JHEP {\bf 1301}, 163 (2013).
  %[arXiv:1210.1267 [hep-ph]].
  %%CITATION = ARXIV:1210.1267;%%
  %6 citations counted in INSPIRE as of 18 Aug 2013
%
%
\bibitem{Miller:2011du} 
  G.~A.~Miller, M.~Strikman and C.~Weiss,
  %``Realizing vector meson dominance with transverse charge densities,''
  Phys.\ Rev.\ C {\bf 84}, 045205 (2011).
  %[arXiv:1105.6364 [hep-ph]].
  %%CITATION = ARXIV:1105.6364;%%
%
%
\bibitem{Schweitzer:2010tt}
  P.~Schweitzer, T.~Teckentrup, A.~Metz,
  %``Intrinsic transverse parton momenta in deeply inelastic reactions,''
  Phys.\ Rev.\  D {\bf 81}, 094019 (2010).
  %[arXiv:1003.2190 [hep-ph]].
%
%
\bibitem{Diakonov:2002fq}
  D.~Diakonov,
  %``Instantons at work,''
  Prog.\ Part.\ Nucl.\ Phys.\  {\bf 51}, 173 (2003).
  %[arXiv:hep-ph/0212026].
  %%CITATION = HEP-PH 0212026;%%
  %%Cited 71 time in SPIRES-HEP
%
%
\bibitem{Diakonov:1987ty}
  D.~Diakonov, V.~Yu.~Petrov and P.~V.~Pobylitsa,
  %``A Chiral Theory Of Nucleons,''
  Nucl.\ Phys.\ B {\bf 306}, 809 (1988).
  %%CITATION = NUPHA,B306,809;%%
  %%Cited 249 times in SPIRES-HEP
%
%
\bibitem{Diakonov:1996sr}
  D.~Diakonov, V.~Petrov, P.~Pobylitsa, M.~V.~Polyakov and C.~Weiss,
  %``Nucleon parton distributions at low normalization point in the large  N(c)
  %limit,''
  Nucl.\ Phys.\  B {\bf 480}, 341 (1996).
  %[arXiv:hep-ph/9606314].
  %%CITATION = NUPHA,B480,341;%%
%
%
\bibitem{Diakonov:1997vc}
  D.~Diakonov, V.~Petrov, P.~Pobylitsa, M.~Polyakov and C.~Weiss,
  %``Unpolarized and polarized quark distributions in the
  %large-N(c) limit,''
  Phys.\ Rev.\ D {\bf 56}, 4069 (1997).
  %[arXiv:hep-ph/9703420].
  %%CITATION = HEP-PH 9703420;%%
  %%Cited 55 times in SPIRES-HEP
%
%
\bibitem{Ji:2013dva}
  X.~Ji,
  %``Parton Physics on Euclidean Lattice,''
  arXiv:1305.1539 [hep-ph].
  %%CITATION = ARXIV:1305.1539;%%
  %4 citations counted in INSPIRE as of 05 Sep 2013
%
%
\bibitem{Diakonov:1985eg}
  D.~Diakonov, V.~Yu.~Petrov,
  %``A Theory of Light Quarks in the Instanton Vacuum,''
  Nucl.\ Phys.\  B {\bf 272}, 457 (1986).
% 
% 
\bibitem{Diakonov:1995qy} 
  D.~Diakonov, M.~V.~Polyakov and C.~Weiss,
  %``Hadronic matrix elements of gluon operators in the instanton vacuum,''
  Nucl.\ Phys.\ B {\bf 461}, 539 (1996).
  %[hep-ph/9510232].
  %%CITATION = HEP-PH/9510232;%%
%
%
\bibitem{Gluck:2007ck}
  M.~Gluck, P.~Jimenez-Delgado and E.~Reya,
  %``Dynamical parton distributions of the nucleon and very small-x physics,''
  Eur.\ Phys.\ J.\  C {\bf 53}, 355 (2008).
  %[arXiv:0709.0614 [hep-ph]].
  %%CITATION = EPHJA,C53,355;%%
%
%
\bibitem{Pobylitsa:1998tk} 
  P.~V.~Pobylitsa, M.~V.~Polyakov, K.~Goeke, T.~Watabe and C.~Weiss,
  %``Isovector unpolarized quark distribution in the nucleon in 
  %the large N(c) limit,''
  Phys.\ Rev.\ D {\bf 59}, 034024 (1999).
  %[hep-ph/9804436].
  %%CITATION = HEP-PH/9804436;%%
%
%
\bibitem{Towell:2001nh}
R.~S.~Towell {\it et al.}  [FNAL E866/NuSea Collab.],
%``Improved measurement of the anti-d/anti-u asymmetry in the nucleon sea,''
Phys.\ Rev.\ D {\bf 64}, 052002 (2001).
%[arXiv:hep-ex/0103030].
%%CITATION = HEP-EX 0103030;%%
%
%
\bibitem{Surrow} B.~Surrow, presentation at QCD Evolution 2013,
Jefferson Lab, May 6--10, 2013.
%
%
\bibitem{Airapetian:2012ki} 
  A.~Airapetian {\it et al.}  [HERMES Collaboration],
  %``Multiplicities of charged pions and kaons from semi-inclusive 
  %deep-inelastic scattering by the proton and the deuteron,''
  Phys.\ Rev.\ D {\bf 87}, 074029 (2013).
  %[arXiv:1212.5407 [hep-ex]].
  %%CITATION = ARXIV:1212.5407;%%
  %5 citations counted in INSPIRE as of 19 Aug 2013
%
%
\bibitem{Adolph:2013stb} 
  C.~Adolph {\it et al.}  [COMPASS Collaboration],
  %``Hadron Transverse Momentum Distributions in Muon Deep 
  %Inelastic Scattering at 160 GeV/$c$,''
  arXiv:1305.7317 [hep-ex].
  %%CITATION = ARXIV:1305.7317;%%
%
%
\bibitem{Accardi:2011mz} 
  A.~Accardi, V.~Guzey, A.~Prokudin and C.~Weiss,
  %``Nuclear physics with a medium-energy Electron-Ion Collider,''
  Eur.\ Phys.\ J.\ A {\bf 48}, 92 (2012).
  %[arXiv:1110.1031 [hep-ph]].
  %%CITATION = ARXIV:1110.1031;%%
%
%
\bibitem{Accardi:2012hwp} 
  A.~Accardi, J.~L.~Albacete, M.~Anselmino, N.~Armesto, 
  E.~C.~Aschenauer, A.~Bacchetta, D.~Boer and W.~Brooks {\it et al.},
  %``Electron Ion Collider: The Next QCD Frontier - 
  %Understanding the glue that binds us all,''
  arXiv:1212.1701 [nucl-ex].
  %%CITATION = ARXIV:1212.1701;%%
  %23 citations counted in INSPIRE as of 04 Sep 2013
%
%
\end{thebibliography}
\end{document}